\documentclass[twocolumn,aps,prl,amssymb,amstext,showpacs]{revtex4} 

\usepackage{epsfig}
\usepackage{amsfonts,amssymb,amsmath}

\newcommand{\beq}{\begin{equation}}
\newcommand{\eeq}{\end{equation}}

\newcommand{\R}{\mathbb{R}}

\newcommand{\m}{{\bf m}}
\newcommand{\field}{{\bf H}}
\newcommand{\xhat}{\hat{{\bf x}}}
\newcommand{\yhat}{\hat{{\bf y}}}
\newcommand{\ud}{\mathrm{d}}

\newcommand{\mt}{{{\partial \m} \over {\partial t}}}

\newcommand{\Dwid}{d_0}

\begin{document}

\title{Domain wall motion in ferromagnetic nanowires driven by arbitrary
  time-dependent fields: An exact result}

\author{Arseni Goussev, JM Robbins, Valeriy Slastikov}

 \affiliation{School of Mathematics, University of Bristol, University
Walk, Bristol BS8 1TW, UK}

\date{\today}

\begin{abstract}
   We address the dynamics of magnetic domain walls in ferromagnetic nanowires under the influence of external time-dependent magnetic fields. We report a new exact spatiotemporal solution of the Landau-Lifshitz-Gilbert equation for the case of soft ferromagnetic wires and nanostructures with uniaxial anisotropy. The solution holds for applied fields with arbitrary strength and time dependence. We further extend this solution to applied fields slowly varying in space and to multiple domain walls.
\end{abstract}


\maketitle


\emph{Introduction.---} The motion of magnetic domain walls (DWs)
in ferromagnetic nanowires has recently become a subject of
intensive research in the condensed matter physics community
\cite{Science-Nature}. Manipulation of DWs by external magnetic
fields, and in particular, the question of how the DW propagation
velocity depends on the applied field have drawn considerable
attention \cite{SunSchliemann10, WangYanLu09, Hickey08}.

In ferromagnetic nanowires, the dynamics of the orientation 
 of the magnetization distribution, $\m(x,t)$ (normalized so that $|\m| = 1$), is described
by the Landau-Lifshitz-Gilbert (LLG) equation \cite{Gilbert55}
\begin{equation}
  \mt + \alpha \m \times \mt = (1+ \alpha^2) \m \times \big( \field(\m)
  + \field_a \big) \,,
\label{eq:LLG}
\end{equation}
where $x$ is the coordinate along the nanowire, $t$ is time,
$\alpha$ is the Gilbert damping parameter, $\field_a$ denotes the
applied magnetic field, and $\field(\m) = -\delta E / \delta \m$,
where
\begin{align}
  E(\m) = {A\over 2} \int_{\R} \left|\frac{\partial \m}{\partial
      x}\right|^2 \! \ud x &+ {K_1 \over 2} \int_{\R} \left(1-
    (\m\cdot\xhat)^2\right) \ud x \nonumber\\ &+ {K_2 \over 2} \int_\R
  (\m\cdot\yhat)^2 \, \ud x.
\label{eq:energy}
\end{align}
is the reduced micromagnetic energy. Here,
$A$ is the exchange constant of the material, and $K_1, K_2 \ge 0$
are the anisotropy constants along the (easy) $x$- and (hard)
$y$-axes. The anisotropy constant along the $z$-axis is taken to
be zero by convention.

To date only one {\it exact} spatiotemporal \footnote{The only other
  exact solution of the LLG equation reported in the literature
  [Z.~Z.~Sun and X.~R.~Wang, Phys. Rev. Lett. {\bf 97}, 077205 (2006)]
  appears in the problem of magnetization switching, where the
  magnetization density is considered to be uniform in space and is a
  function of time only, i.e., $\m=\m(t)$.} solution of the LLG
equation has been reported in the literature, namely the so-called
Walker solution \cite{SchryerWalker74}.  
The analysis of
Schryer and Walker \cite{SchryerWalker74} applies to the case  where $K_2 > 0$,
ie where the anisotropy constants in the transverse plane are strictly unequal.  This
is appropriate for
a thin film or thin strip geometry.
The applied field
is taken to be uniform in space, constant in time, and directed along the
nanowire, i.e., $\field_a(x,t) = H_a \xhat$.
For $|H_a|$ less than a certain threshold $H_W$, the
so-called Walker breakdown field, a planar domain wall propagates
rigidly along the nanostrip with
velocity depending nonlinearly on $H_a$.

%

In this Letter we present an {\it exact} spatiotemporal solution
of the LLG equation that, to our knowledge, has not been
previously reported in the literature. We consider the case of
transverse isotropy, ie $K_2 = 0$.  This is appropriate for soft ferromagnetic
nanowires whose
cross-sectional dimensions are comparable, as well as for uniaxial nanowires
whose easy axis lies along the wire.
We
take the applied field to lie along the nanowire, as in the case of the Walker solution,
but allow for
arbitrary time dependence, i.e., $\field_a(x,t) = H_a(t) \xhat$.


\emph{Exact solution of the LLG equation.---} The boundary
conditions appropriate for a domain wall  with finite micromagnetic energy
$E(\m)$ are given by
$\m(x,t) \rightarrow \pm \xhat$ as $x \rightarrow \pm \infty$.  For
$K_2=0$ the magnetization-dependent field $\field$ is given by
\begin{equation}
  \field(\m) = A\frac{\partial^2 \m}{\partial x^2}
  + K_1 (\m \cdot \xhat) \xhat \,.
\label{eq:effective_field}
\end{equation}
We now take into account the fact that $\m$ has its values on
$S^2$, and parametrize $\m$ in terms of angles $\theta(x,t)$ and
$\phi(x,t)$ according to $\m = (\cos\theta, \sin\theta \cos\phi, \sin\theta
\sin\phi)$.
From Eqs.~\eqref{eq:LLG} and \eqref{eq:effective_field} we obtain
the LLG equation in the equivalent form
\begin{subequations}
\begin{align}
  &\dot{\theta} - \alpha \dot{\phi} \sin \theta + A (1+ \alpha^2)
  \big( \phi'' \sin\theta + 2 \theta' \phi' \cos \theta \big) = 0 \,, \label{eq:angle_eq1}\\
  &\alpha \dot{\theta} + \dot{\phi} \sin\theta + (1+ \alpha^2) \big(
  -A \theta'' + A(\phi')^2 \sin\theta \cos\theta \nonumber\\
  &\phantom{xxxxxxxxxxx} + K_1 \cos\theta \sin\theta + H_a(t) \sin\theta
  \big) = 0 \,, \label{eq:angle_eq2}
\end{align}
\label{eq:angle_eq} \end{subequations} where dot\ \ $\dot\empty$\
\ denotes $\partial/\partial t$ and prime\ \  $'$  \  denotes
$\partial/\partial x$.

We now look for a solution of Eq.~\eqref{eq:angle_eq} in the form
\begin{equation}
  \theta_*(x,t) = \theta_0 \left( x - x_* (t) \right) \,, \quad \phi_*(x,t) = \phi_* (t) \,,
\label{eq:solution}
\end{equation}
where
\begin{equation}
  \theta_0(x) = 2 \arctan \exp \left( -x/\Dwid  \right),\ \ \Dwid = \sqrt{A/K_1}.
\label{eq:profile}
\end{equation}
$\theta_0(x)$ describes the static domain wall  in the absence
of an applied field.
The magnetization density  determined by $\theta_0(x)$ and $\phi_0(x) = \pi/2$ minimizes
the micromagnetic energy $E(\m)$ for the specified boundary conditions.
Substituting
Eq.~\eqref{eq:profile} into Eq.~\eqref{eq:angle_eq}, and taking
into account that $\theta_0' = - \sin\theta_0/\Dwid$ and
$\theta_0'' = \sin 2\theta_0/(2\Dwid^2)$, we find that $\theta_*$ and $\phi_*$
satisfy the LLG equation~\eqref{eq:angle_eq} provided that
$x_*(t)$ and $\phi_*(t)$ satisfy
\begin{equation}
  \dot{x}_* = - \alpha  \Dwid H_a(t)    \,, \quad \dot{\phi}_* = -H_a(t) \,.
\label{eq:answer}
\end{equation}
(In fact, \eqref{eq:profile} and \eqref{eq:answer} provide the only solution of
the form \eqref{eq:solution}.)

Equations~(\ref{eq:solution}-\ref{eq:answer}) constitute the main
result of this Letter. They represent an {\it exact} solution of
the LLG equation, and describe a
DW, with profile independent of the applied field,
propagating along the nanowire with velocity $\dot{x}_*$ while
precessing about the nanowire with angular velocity
$\dot{\phi}_*$. {\it No restrictions have been imposed on the
strength of the applied magnetic field and no assumptions have
been made about its time dependence.}

We now compare the {\it precessing solution}
Eqs.~(\ref{eq:solution}-\ref{eq:answer}) with the Walker solution
\cite{SchryerWalker74}.  The Walker solution is defined only for
$K_2
> 0$ (the fully anisotropic case) and time-independent $H_a$ less than the  breakdown
field
\begin{equation}\label{eq: H_W}
H_W = \alpha K_2/2.
\end{equation}
It is given by
\begin{equation}
  \theta_W(x,t) = \theta_0
   \left(\frac{x - V_Wt}{\gamma}\right)  \,,
  \quad \phi_W(x,t) = \phi_W \,,
\label{eq:Wsolution}
\end{equation}
where
\begin{equation}\label{eq: phi_W}
\sin 2\phi_W =  H_a/H_W
\end{equation}
determines the (fixed) inclination of the DW plane and
\begin{equation}\label{eq: V_W gamma_W}
V_W = \gamma \frac{1+\alpha^2}{\alpha} \Dwid H_a,\ \gamma =
\left(\frac{K_1}{K_1 + K_2\cos^2\phi_W}\right)^{1/2}
\end{equation}
gives the DW velocity.

There are several characteristic differences between the
Walker solution and  the precessing solution which should be distinguishable experimentally.
Foremost is the fact that the Walker
solution exists only for constant applied fields whose strength does not exceed
a certain threshold, so that
the DW velocity is bounded.
The precessing solution is defined for time-dependent
applied fields of arbitrary strength, so that the DW velocity, which for the precessing solution is proportional to the field
strength, can be arbitrarily large.
Next, while for the Walker solution
the plane of the DW remains fixed, for
the precessing solution
it rotates about the nanowire at a rate proportional to $H_a$.
Finally, we observe that,  for the Walker solution,  the DW profile
contracts ($\gamma > 1$) or expands ($\gamma > 1$) in response to
the applied field, 
whereas for the precessing
solution the DW profile propagates without distortion.


\emph{Spatially nonuniform applied fields and multiple domain
  walls.---} We now extend our results to applied fields that depend
on both position along the nanowire and time, i.e, $\field_a =
H_a(x,t) \xhat$. For any (non-singular) applied field,
Eq.~\eqref{eq:angle_eq} is satisfied at $x$ outside the DW
transition layer $|x-x_*(t)| \gg \Dwid$ (up to exponentially small terms). Assuming now that the
field varies slowly across the transition region,
\begin{align}
  \big| H_a\big(x,t\big)&-H_a\big(x_*(t),t\big) \big| \ll \big|
  H_a\big(x_*(t),t\big) \big| \nonumber\\ &\mathrm{for} \quad
  |x-x_*(t)| \lesssim \Dwid \,,
\label{eq:slow_field}
\end{align}
we obtain an approximate solution of the LLG equation: the
magnetization density is given by Eqs.~\eqref{eq:solution} and
\eqref{eq:profile} with
\begin{equation}
  \dot{x}_* = - \alpha \Dwid H_a\big(x_*(t),t\big)
  \,, \quad \dot{\phi}_* = -H_a\big(x_*(t),t\big) \,.
\label{eq:answer_approx}
\end{equation}
The physical meaning of Eq.~\eqref{eq:answer_approx} is quite obvious:
the DW is only sensitive to the applied field within the
transition layer.

\begin{figure}[h]
\centerline{\epsfig{figure=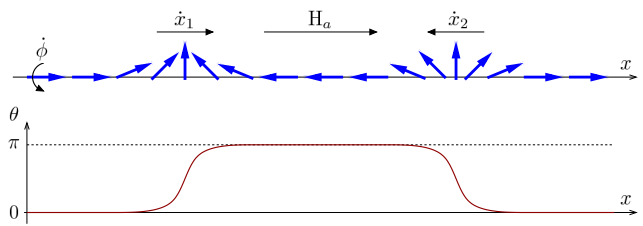,width=3.3in}}
\caption{(Color online) Dynamics of domain walls. See text for discussion. }
\label{fig1}
\end{figure}

This approximation can now be extended to the case of $N$
non-overlapping DWs. Indeed,
\begin{subequations}
\begin{align}
  &\theta_N(x,t) = \sum_{n=1}^N \theta_0 \big( (-1)^{n+1} (x - x_n (t))
  \big) \,, \label{eq:multiple_dw1}\\ &\phi_N(x,t) = \phi_{\bar{n}}(t)
  \,, \quad n=\bar{n} \; \mathrm{minimizes} \; |x-x_n(t)|
  \,, \label{eq:multiple_dw2}
\end{align}
\label{eq:multiple_dw}
\end{subequations}
with $x_{k+1}(t)-x_k(t) \gg \Dwid$ for $k=1,\ldots,N-1$,
constitutes an approximate solution of the LLG equation given that
\begin{subequations}
\begin{align}
  &\dot{x}_n = (-1)^n \alpha \Dwid H_a\big(x_n(t),t\big)
  \,, \label{eq:multiple_solutions1}\\ &\dot{\phi}_n = -
  H_a\big(x_n(t),t\big) \,, \label{eq:multiple_solutions2}
\end{align}
\label{eq:multiple_solutions}
\end{subequations}
for $n=1,\ldots,N$. For the case of a spatially uniform applied field
Eqs.~\eqref{eq:multiple_dw} and \eqref{eq:multiple_solutions} describe
the time evolution of $N$ DWs such that any two adjacent WDs travel in
opposite directions while rotating in the same direction (and with the
same angular velocity) around the nanowire.


\emph{Conclusions.---} In this Letter we have presented an exact
spatiotemporal solution of the LLG equation that has not been
previously reported in the literature. The validity of the new
solution requires no assumptions about the time-dependence or strength
of the applied field.

We have then provided a natural extension of the solution to physical
situations in which the applied field varies slowly in space. 
An approximate solution of the LLG equation for the case
of multiple domain walls has also been obtained.


\emph{Acknowledgments.--} A.G. acknowledges the support by EPSRC under
Grant No. EP/E024629/1.


\end{document}